\documentclass{aastex}          
\usepackage{spr-astr-addons}    

\begin{document}
%
\title{Solar Influence on Nuclear Decay Rates: Constraints from the MESSENGER Mission}

\shorttitle{MESSENGER Mission Constraints on $^{137}$Cs Decays}
\shortauthors{Fischbach et al.}

\author{E. Fischbach\altaffilmark{1}}
\affil{Department of Physics, Purdue University, West Lafayette, IN 47907 USA}
\and
\author{K.J. Chen}
\affil{Department of Mathematics, Purdue University, West Lafayette, IN  47907 USA}
\author{R.E. Gold} \and \author{J.O. Goldsten} \and \author{D.J. Lawrence} \and \author{R. J. McNutt, Jr.} \and \author{E.A. Rhodes}
\affil{Johns Hopkins University Applied Physics Laboratory, Laurel, MD  20723-6099 USA}
\and
\author{J.H. Jenkins\altaffilmark{2}}
\affil{School of Nuclear Engineering, Purdue University, West Lafayette, IN  47907 USA}
\and
\author{J. Longuski}
\affil{School of Aeronautical and Astronautical Engineering, Purdue University, West Lafayette, IN  47907 USA}

\altaffiltext{1}{525 Northwestern Ave. West Lafayette, IN 47907 USA, ephraim@purdue.edu}
\altaffiltext{2}{Department of Physics, Purdue University, West Lafayette, IN  47907 USA}

\begin{abstract}
We have analyzed $^{137}$Cs decay data, obtained from
a small sample onboard the MESSENGER spacecraft en route to Mercury, with the aim of setting
limits on a possible correlation between nuclear decay
rates and solar activity.  Such a correlation has been
suggested recently on the basis of data from $^{54}$Mn decay
during the solar flare of 13 December 2006, and by
indications of an annual and other periodic variations in the decay
rates of $^{32}$Si, $^{36}$Cl, and $^{226}$Ra.
Data from five measurements of the $^{137}$Cs count rate over a period of approximately 5.4 years
have been fit to a formula which accounts for the usual exponential decrease in count rate
over time, along with the addition of a theoretical solar contribution varying with MESSENGER-Sun
separation. The indication of solar influence is then characterized by a non-zero value of the calculated parameter $\xi$,
and we find $\xi=(2.8\pm8.1)\times10^{-3}$ for $^{137}$Cs. A simulation of the increased data
that can hypothetically be expected following Mercury orbit insertion on 18 March 2011 suggests that
the anticipated improvement in the determination of $\xi$ could reveal a non-zero value
of $\xi$ if present at a level consistent with other data.
\end{abstract}

\keywords{astroparticle physics -- nuclear reactions -- Sun:particle emission}

In a recent series of papers \citep{jen09a,jen09b,fis09,stu10a,jav10,stu10b} evidence has been
presented for a possible solar influence on nuclear decay
rates.  Data analyzed by \citet{jen09a} and \citet{fis09} indicate a possible
correlation between the solar flare of 13 December 2006
and a decrease in the measured decay rate of $^{54}$Mn coincident in
time with the flare.  An analysis of data from
Brookhaven National Laboratory (BNL) on the measured decay rates of $^{32}$Si and
$^{36}$Cl, and from the Physikalisch-Technische Bundesanstalt
(PTB) in Germany on the measured decay rates of $^{226}$Ra and its daughters, show that both data sets exhibit
similar annual variations in their respective decay rates \citep{jen09b,fis09}.
Similar periodic effects have been reported by \citet{par04,par05,par10b,par10c,par10d}, \citet{ell90},  \citet{fal01}, \citet{Baur01,Baur07}, and citet{shn98,shn00}, and more recently by \citet{jen11}, in data from The Ohio State University. In addition to annual periodicities, evidence for other periodicities in decay data possibly associated with solar rotation is reported in 
\citet{stu10a}, \citet{stu10b}, \citet{fis10}, and \citet{stu11}, including evidence for a period of $\sim$33 days, and for a 2.11 yr$^{-1}$ Rieger-like
periodicity. Since none of the rotation-related periodic signals (in what should be randomly distributed data) corresponds to any known terrestrial
influence, these results support the inference of a solar origin to time-varying nuclear decay rates,
through some as yet unknown mechanism \citep{fis11}.

The interpretation of the data in \citet{jen09a}, \citet{jen09b}, and \citet{fis09} have been questioned by \citet{coo09,nor09,sem09}, however, \citet{jen10} addressed all of those questions.
Nonetheless, the nature of the effects
reported in the original references remains uncertain at
present, particularly since solar flares are not always correlated in detectable changes in nuclear decay rates \citep{par10a}.  However, if nuclear decays can in fact be
influenced by the Sun, and specifically by the varying distance between a decaying source
and the Sun, then data from the MErcury Surface, Space ENvironment,
GEochemistry, and Ranging (MESSENGER)
mission to Mercury \citep{sol07} could lead to significant
constraints on the magnitudes of such effects on the examined isotope, given that
the MESSENGER-Sun distance has varied from 1.0689 AU to
0.30748 AU. The object of the present paper is to use
decay data from a sample of $^{137}$Cs onboard MESSENGER
to search for a possible variation of the $^{137}$Cs decay
parameter $\lambda$, or its half-life $T_{1/2} = \ell n 2/\lambda =
30.07$ yr \citep{Bau02}, over the course of the mission to date.

\begin{figure}[h]
\includegraphics[width=\columnwidth]{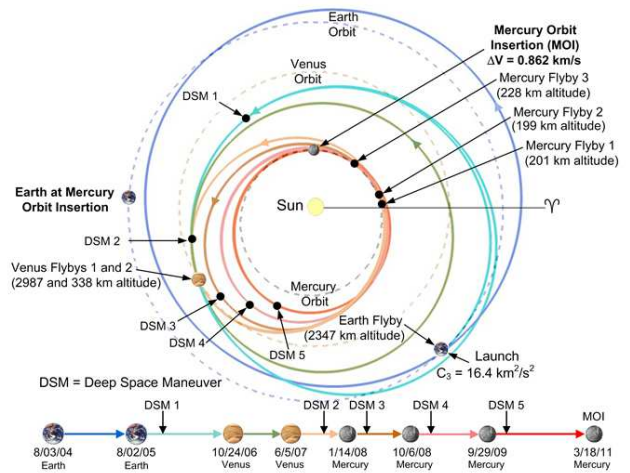}
\caption{Timeline for the MESSENGER mission. Launch from Earth was on 4 March 2004, and Mercury
orbit insertion (MOI) occurred on 18 March 2011. During the MESSENGER mission the MESSENGER-Sun distance has varied from
1.0689 A.U. to 0.30748 A.U., a wide enough range to test the hypothesis of a time-varying nuclear decay rate
that depends on distance from the Sun. Figure courtesy of NASA/Johns Hopkins University Applied Physics
Laboratory/Carnegie Institution of Washington. \label{fig:mission}}
\end{figure}

The MESSENGER spacecraft was launched on 3 August 2004,
and Mercury orbit insertion (MOI) occurred as scheduled on 18 March 2011.
The objective of MESSENGER is to study Mercury from a highly eccentric orbit with 
a low-altitude periapsis. Following launch, several gravity-assist maneuvers 
were carried out by the MESSENGER spacecraft. These included a
close flyby of the Earth approximately one year after launch (to reduce
the required launch energy), and a subsequent deep-space maneuver
to put MESSENGER on course for two Venus flybys to further lower the
perihelion distance. Three Mercury flybys (with a leveraging deep-space
maneuver in between successive Mercury gravity assists) were used to gradually
slow the spacecraft with respect to Mercury. (A timeline for the MESSENGER
mission trajectory is presented in Figure \ref{fig:mission}.)
Among the seven instruments that comprise the MESSENGER
physics payload is a Gamma-Ray and Neutron Spectrometer (GRNS) \citep{gol07}, whose
purpose is to map the elemental composition of the
surface of Mercury. Galactic cosmic rays (GCRs) incident on the surface of 
Mercury produce characteristic gamma-rays and neutrons, which are detected 
by the GRNS, from which inferences can be drawn about the planet's surface 
composition. The Gamma-Ray Spectrometer (GRS) portion of the GRNS is a 
high-purity germanium (HPGe) sensor surrounded by a plastic scintillator 
anticoincidence shield (ACS).  During normal operation, the HPGe sensor is 
cooled to cryogenic temperatures using a miniature mechanical cooler.  The mechanical 
cooler has a limited operational life, with an expected mean time to failure of 
approximately one year.  Because the primary 
goal of the GRS is to map Mercury's composition for one Earth year, the GRS has been 
used sparingly during MESSENGER's seven-year cruise phase.  The few times the GRS has 
been turned on have been to verify its operation and to make gamma-ray measurements of 
Mercury during the three Mercury flybys \citep{rho11}.  

Shortly after launch,
readings from the GRS while in deep space revealed the presence of an unexpected
small $^{137}$Cs source. It is not known if this $^{137}$Cs source is incorporated into
the GRS sensor or some part of its housing. Nevertheless, because this source has
been onboard MESSENGER
from launch, its presence allows us, serendipitously, to set limits on the influence of solar activity
on $^{137}$Cs decay. To date, five sets of measurements of the $^{137}$Cs decay rate 
have been made during periods denoted by Cruise 0 through Cruise 4, and more are expected
after MOI. In what follows, we develop a formalism to study a possible solar influence
on $^{137}$Cs decay, and we then apply this formalism to the existing data. We conclude with a
discussion of the improvements that can be expected from data to be acquired after MOI.

\section{Theoretical Formalism}

The presence of the $^{137}$Cs source onboard MESSENGER provides an opportunity to
test whether the $^{137}$Cs decay rate varies with 
MESSENGER-Sun distance.
Following the discussion presented in \citet{fis09} we assume that the
beta decay rate $dN(t)/dt \equiv \dot{N}(t)$ of a sample
containing $N(t)$ unstable nuclei can be written in the
form 
\begin{eqnarray}
  \frac{-dN(t)}{dt} = \lambda(t)N(t) =
      [\lambda_0 + \lambda_1(\vec{r},t)]N(t),
      \label{Decay}   
\end{eqnarray}
\noindent{}where $\lambda_0$ represents the intrinsic contribution
to the $\beta$-decay rate arising from the conventional
weak interaction along with a possible time-independent
background arising from new interactions. The time-dependent perturbation
$\lambda_1(\vec{r},t)$ is also a function of the distance
$r = |\vec{r}|$ between the decaying nucleus and the source,
here assumed to be the Sun.  For a perturbation whose
influence on a decaying nucleus varies as $1/r^n$, where $n$ is an integer,
$\lambda_1(\vec{r},t)$ can be expressed in the form
\begin{eqnarray}
  \lambda_1(\vec{r},t) = \lambda_0\xi^{(n)} \left[ \frac{R}{r(t)} \right]^n ,
      \label{lambda1}   
\end{eqnarray}
\noindent
where $R \equiv r(t=0)$, and $\xi^{(n)}$ is the parameter we are
interested in constraining. In what follows we focus initially on $\xi^{(2)}\equiv\xi$, which 
is the most likely variation that we expect, (i.e., an inverse-square law). Results for $n=1$ and $n=3$, which are the next most
likely alternatives to $n=2$, can be analyzed in a similar way. We note from Eqs. (\ref{Decay}) and (\ref{lambda1})
that $\xi$ is not a universal constant since it depends
on the specified initial values $R$ and $t=0$.  In the
present paper we define $t=0$ as the launch time,
in which case $R = 1~\textrm{AU} = 1.495979 \times 10^{8}$ km.
Integrating Eq. (\ref{Decay}) we find
\begin{eqnarray}
  N(t) = N_0 \exp \{ -\lambda_0 [t + \xi I(t)]\},
      \label{Noft}   
\end{eqnarray}
\begin{eqnarray}
\nonumber  -\frac{dN(t)}{dt} = \\ \lambda_0N_0 \exp\left(-\lambda_0t\right)
     \left\lbrace 1 + \xi \left[ 
        \frac{R^2}{r^2(t)} - 
        \lambda_0 I(t)\right]\right\rbrace, 
\label{fulldecay} 
\end{eqnarray}
\begin{eqnarray}
  I(t) = \int_0^t dt^\prime \left[ \frac{R}{r(t')}\right]^2.
\label{I2} 
\end{eqnarray} 
\noindent We note from Eqs. (\ref{Noft}-\ref{I2}) that since $I(0) = 0$,
the effective $^{137}$Cs decay constant at
$t = 0$ is $\lambda \equiv \lambda_0(1 + \xi)$,
where $\lambda = 6.311(19) \times 10^{-5}~\textrm{d}^{-1}$
is the standard value measured on Earth (at 1 AU)
corresponding to $T_{1/2} = 30.07(9)$ yr.
It is convenient to rewrite Eq. (\ref{fulldecay}) in terms of
directly measured quantities by first expressing
$\lambda_0$ in terms of $\lambda$, and then eliminating
the unknown $N_0$ by taking appropriate ratios.
To lowest order in $\xi \ll 1$ we find 
\begin{eqnarray}
\nonumber  e^{+\lambda t} \frac{dN(t)/dt}{dN(t=0)/dt}
    &\cong& 1 + \xi \left[\frac{R^2}{r^2(t)}
    - \lambda I(t) + \lambda t-1\right] \\
     &\equiv& 1 + \xi B(t).
\label{Boft}   
\end{eqnarray}
\noindent
We note from Eq. (\ref{Boft}) that since $B(t=0) = 0$ the count rate ratio approaches unity as $t \rightarrow 0$,
as expected. Comparing Eqs. (\ref{fulldecay}) and (\ref{Boft}), the additional terms in the square brackets in Eq. (\ref{Boft}) arise from
computing the indicated ratio, and then expressing the final result in terms of the laboratory value $\lambda$.

\begin{figure}[h]
\includegraphics[width=\columnwidth]{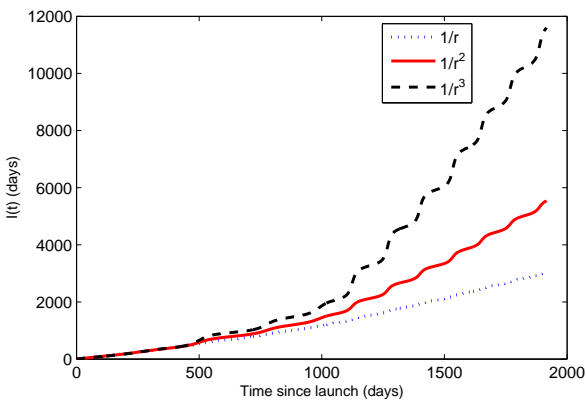}
\caption{Plots of the integrals $I(t)$, for $n=1,2,3,$ in
Eq. (\ref{lambda1}).  The integrals are defined in analogy to $I^{(2)}(t)$
in Eq. (\ref{I2}). \label{fig:Rcurves}}
\end{figure}

In principle, the left hand side of Eq. (\ref{Boft}) and $B(t)$ are
directly measurable, and hence can be used to set limits
on $\xi$. These limits can then be compared to those
obtained in the laboratory.  In practice, however,
$dN(t=0)/dt$ was not measured, and hence our limits on $\xi$
were determined from ratios of count rates obtained from
measurements made during the five cruise periods. A summary of the data obtained
from these measurements is presented in Table \ref{Tbl:Detail}. Among
the entries in this table, the precision of the data which
enter into the calculation of $B(t)$ is such that the errors
on $B(t)$ can be neglected compared to those arising from the
count rates (see discussion below). The integral $I(t)$, which is
exhibited in Figure \ref{fig:Rcurves}, accounts for the cumulative $\xi$-dependent
contributions to the $^{137}$Cs decay rate arising from both
the change in MESSENGER-Sun distance, as well as from the
normalization relation $\lambda = \lambda_0(1 + \xi)$.  

\begin{table*}
\footnotesize
\caption{Summary of $^{137}$Cs decay data from the MESSENGER mission. For each of the five measurement periods ($t_0,t_1,t_2,t_3,t_4$), these data include the calculated values of $I(t)$, $B(t)$, and $\Delta(t)$, for a perturbation varying as $1/r^2$\label{Tbl:Detail}}
\begin{tabular}{@{}lrrrrr@{}}
\tableline
  & Cruise 0 & Cruise 1 & Cruise 2 & Cruise 3 & Cruise 4 \\
&  Initial checkout & Mercury Flyby 1 & Mercury Flyby 2 & Mercury Flyby 3 & Final checkout \\  
\tableline
Calendar date & 16 Nov 2004 & 14 Jan 2008 & 6 Oct 2008 & 2 Oct 2009 & 14 Apr 2010 \\
Julian date  &  2453322.500  &  2454478.500  &  2454744.500  &  2455102.500 & 2455300.500 \\
Distance from the Sun (km)  &  160599579.0277  &  54595403.8012  &  52751315.2631  &  48132083.3548 & 63523375.7902 \\
Elapsed time since Cruise 0 (days)  &  0.0  &  1156.0  &  1422.0  &  1780.0 & 1978.0 \\
Mean count rate in peak (cps) &  0.100$\pm$0.0012  &  0.0942$\pm$0.0019  &  0.0933$\pm$0.0020  &  0.0886$\pm$0.0017 & 0.0886$\pm$0.0020\\
Total accumulation time (hours) & 60.00 &  45.00  &  40.00  &  60.00  &  39.00 \\
$I(t)$ (days)  &  91.173  &  2284.190  &  3408.077  &  5162.823 & 6228.370  \\
$B(t)$  &  -0.1316  &  6.4397  &  6.9235  &  8.4530 & 4.2842 \\
$\Delta(t)$  &  0.0000  &  6.5751  &  7.0551  &  8.5847 & 4.4159 \\
\tableline
\end{tabular}

\end{table*}

Returning to Eq. (\ref{Boft}), the experimentally interesting quantities are
the ratios
\begin{eqnarray}
\nonumber \frac{[\exp(\lambda t_i)dN(t_i)/dt]}{[\exp (\lambda t_0)dN(t_0)/dt]},
\end{eqnarray}
\noindent
where $t_0$ and $t_i$ $(i = 1,2,3,4)$ denote the start times
of the five cruise measurement periods.  These five measurements
thus determine four ratios which, along with the corresponding
values of $B(t_i$), can be used to obtain $\xi$.  

\section{Analysis of the $^{137}$Cs Count Rates}

The MESSENGER GRS data analyzed in this paper are made publicly available in the NASA Planetary 
Data System (PDS) archive\footnote{(http://geo.pds.nasa.gov/missions/messenger/index.htm)} according to 
the schedule given at the PDS web site.  Table \ref{Tbl:Detail} lists the time periods for each of the five cruise 
measurements.  Each of the data collection periods lasted for greater than 24 hours, and some for 
over 60 hours.  To ensure that the data analyzed here were acquired when the instrument parameters and 
background were reasonably constant, appropriate time windows were selected within each data 
collection period. For example, the Cruise 0 
period took place during the declining phase of a solar particle event, so the background was 
relatively high during the early portion of this collection period.  Thus, the selected time window 
for the Cruise 0 period was the last 60 hours of the collection period, when the background was 
the lowest.  For the other data collection periods, data were excluded during the Mercury flybys 
and when instrument parameters were varied for sensor testing.

\begin{figure}[h]
\includegraphics[width=\columnwidth]{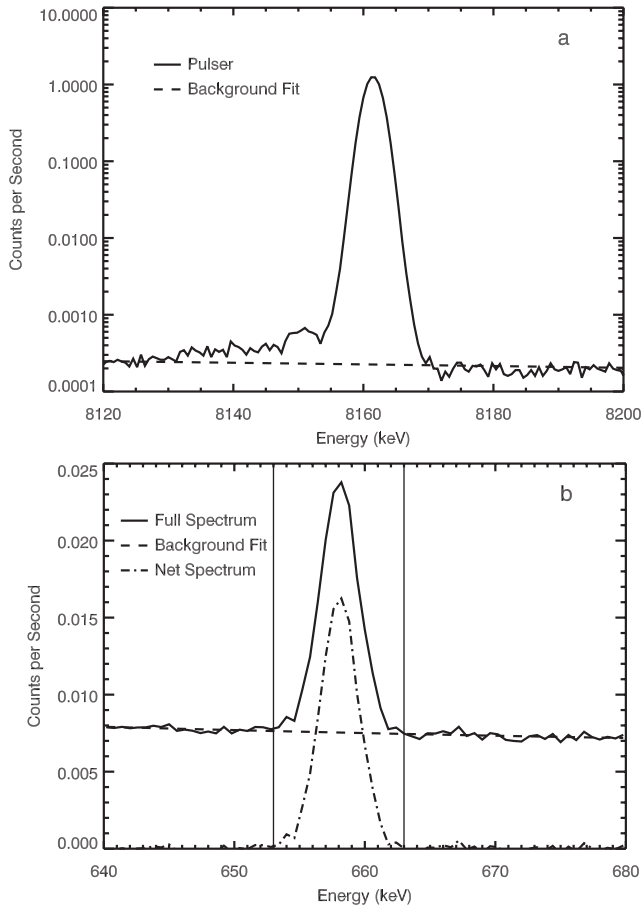}
\caption{Measured GRS pulse height spectra taken during the Cruise 0 data collection 
period.  (a) Pulser data (solid line) and background (dashed line) fit to the pulser 
data; (b) $^{137}$Cs data (solid line), background in the region around 
the $^{137}$Cs line (dashed line), and net $^{137}$Cs spectrum (dot-dashed line).  Solid 
vertical lines show the energy window over which the counts are determined. \label{fig:peak}}
\end{figure}

Figure \ref{fig:peak} shows an example anticoincidence pulse-height spectrum from the Cruise 0 data 
collection period.  Figure \ref{fig:peak}a shows the electronic pulser peak used to correct for detector 
dead time; Figure \ref{fig:peak}b shows the 662 keV peak from the $^{137}$Cs decay.  GRS dead time is largely 
driven by a ~1000 Hz raw counting rate in the ACS.  The dead time is monitored by 
a 7.62939 Hz electronic pulse fed directly across the HPGe detector.  The pulser line appears 
at a high energy in the pulse-height spectrum, far away from other prominent gamma-ray 
lines.  The total counts in the pulser peak are obtained by fitting a linear function to 
the underlying background, subtracting this background from the pulse height spectra, and 
summing the net counts.  The live time fraction is the ratio of the measured counts to the 
expected number of pulser counts.  For the five measurement periods, the live time fraction 
varied from 0.9 to 0.94. 

The total counts in the $^{137}$Cs peak are determined by dividing the spectra by the live time 
fraction, fitting a second-degree polynomial function to the background above and below the 
peak (dashed line in Figure \ref{fig:peak}b), subtracting this background, and then summing the net 
counts (dot-dashed line) within a window around the peak.  The count rate uncertainties 
are calculated by assuming all uncertainties are due to Poisson counting 
statistics.  Specifically, if the total net counts within the energy window 
are $N_{Net}=N_{Spec}-N_{Back}$, where $ N_{Spec} $ and $ N_{Back} $ are the 
total spectrum and background counts within the energy window, respectively, then the total 
count uncertainty, $ \sigma_{Net} $, is

\begin{eqnarray}
\sigma_{Net}=\sqrt{\sigma^2_{Spec}+\sigma^2_{Back}}=\sqrt{N_{Spec}+N_{Back}}.
\label{uncertainty}   
\end{eqnarray}

\noindent{}The corresponding count rate uncertainty is:

\begin{eqnarray}
\sigma_{dN/dt}=\left( \frac{dN}{dt}\right) \frac{\sigma_{Net}}{N_{Net}}.
\label{uncertaintycr}   
\end{eqnarray}

\noindent{}The uncertainty associated with the live time correction is neglected 
because its magnitude is almost an order of magnitude lower than that of the 
uncertainty for the $^{137}$Cs peak.

One other point regarding the GRS is that in order to reduce energy spreading effects from GCR 
radiation damage, the GRS is periodically annealed at high temperatures.  Since launch, the GRS 
has been annealed four times (no annealing was carried out prior to Cruise 4).  One consequence of annealing is that 
the active volume of the germanium crystal may shrink slightly, which in turn will slightly 
reduce the overall detector efficiency. By monitoring the position of the pulser peak after 
each annealing operation, we can estimate the change in detector capacitance, and therefore 
the active detector volume, by assuming a simple coaxial geometry for the HPGe detector. Since 
launch, the pulser peak has shifted $ \sim $7\%, which equates to less than $ \sim $0.7\% loss in detector 
volume; however, the actual change in detector efficiency at 662 keV is not easily 
calculated (perhaps it could be better modeled through a Monte Carlo simulation), but 
would likely lie between the estimates for the capacitance and volume. Such efficiency 
changes have not been included in this analysis.  For future analyses, efficiency 
corrections can be monitored and/or included by observing changes in peak counts 
associated in time with annealing events, as well as by modelling the GRS efficiency response.

\section{Results}

Following the preceding discussion, we have determined $\xi$
by using Eq. (\ref{Boft}) to write
\begin{eqnarray}
    e^{\lambda(t_i-t_0)}
    \frac{[dN(t_i)/dt]}
         {[dN(t_0)/dt]} \equiv \xi \Delta(t_i) + 1,
\label{Delta1}
\end{eqnarray}
\noindent{}where $i=1,2,3,4$ and $\Delta(t_i)=B(t_i)-B(t_0)=B(t_i)+0.1316$. The starting dates $t_i$ 
for Cruise 1, \ldots{} Cruise 4, are given in Table \ref{Tbl:Detail},
as are the calculated values for the integrals $I(t_i)$
in Eq. (\ref{I2}), and $\Delta(t_i)$ in Eq. (\ref{Delta1}).  
Combining these results with the four ratios obtained from the measured
countrates allows us to fit Eq. (\ref{Delta1}) to a straight line,
$y_i = \xi x_i + b$, whose slope is $\xi$.  We find,
\begin{eqnarray}
    \xi \equiv \xi^{(2)}= (2.8\pm{}8.1)\times10^{-3};~~~b^{(2)}=1.02\pm0.03
\label{Delta2}
\end{eqnarray}

The result for $\xi$ in Eq. (\ref{Delta2}) can be compared to the annual variation in 
the $^{32}$Si/$^{36}$Cl count rate ratio
reported by \citet{jen09a} from their analysis of the data of \citet{alb86}
taken at BNL. \citet{jen09a} found a fractional change of $\sim3\times10^{-3}$
in $\dot{N}(^{32}{\rm Si})/\dot{N}(^{36}{\rm Cl})$ from perihelion to 
aphelion, and this level of sensitivity may be achievable following 
MOI given the result in Eq. (\ref{Delta2}) as we now discuss.

\section{Post-MOI Simulation}

The result in Eq. (\ref{Delta2}) suggests that
the availability of a much larger data set following MOI on 18 March 2011 could lead to significantly
improved constraints on $\xi$ for $^{137}$Cs. In this section, we simulate post-MOI data that can be expected in order to estimate
the level of improvement on $\xi$ that might be
realized.

We can estimate the post-MOI countrates by re-expressing Eq. (\ref{Boft}) and (\ref{Delta1}) in the form
\begin{eqnarray}
\nonumber e^{\lambda\Delta{}t}\left[\frac{dN(t_i)/dt}{dN(t_1)/dt}\right]
    &\cong&  \\ 1  + \xi \left[\frac{R^2}{r^2(t_i)}
    - \lambda \int_{t_1}^{t_i}dt^\prime \left[ \frac{R}{r(t')}\right]^2 
    + \lambda\Delta t-1-D\right],
\label{SimRat}   
\end{eqnarray}
\noindent{}where
\begin{eqnarray}
D=\frac{R^2}{r^2(t_1)} - \lambda \int_{0}^{t_1}dt^\prime \left[ \frac{R}{r(t')}\right]^2 + \lambda t_1 -1 = 9.1542,
\label{DDef}
\end{eqnarray}


accounts for the accumulated change in the $^{137}$Cs count rate between Earth launch at $t=0$ and MOI at $t=t_1$.
In Eq. (\ref{SimRat}), $t_i$ is the time of the $i^{th}$ measurement, $\Delta t=t_i-t_1$ in days since MOI, and we
have used for $r\left( t\right)$ the actual (planned) trajectory ephemeris, calculated by the 
Jet Propulsion Laboratory HORIZONS system. The simulation was run for slightly more than one Earth-year
following MOI, during which each measurement was assumed to last for 11 days. We have also assumed that $\xi=-0.0031$,
which is compatible with the data of \citet{alb86}, and have fixed the initial $^{137}$Cs count rate
to be 0.08 cps, which is suggested by the data in Table \ref{Tbl:Detail}. Under these assumptions, the standard fractional
statistical error in each 11-day run is approximately $\Delta \dot{N_i}/\dot{N_i}=0.0036$, and hence the fractional
error in the ratio given in Eq. (\ref{SimRat}) is approximately $\sqrt{2}(0.0036)=0.0051$. A uniformly distributed
random error was then added to each expected measurement to simulate uncertainties in the observations. The objective
of the simulation is to recover the input value $\xi=-0.0031$, and to then determine its associated uncertainty.

\begin{figure}[h]
\includegraphics[width=\columnwidth]{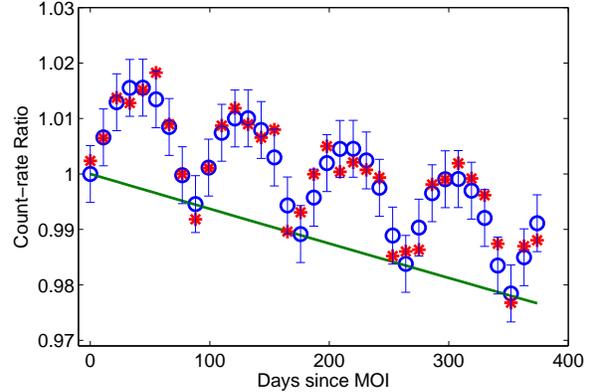}
\caption{Simulated count rates as a function of time. The error bars represent
the $1/\sqrt{N}$ statistical uncertainties at each point, and the asterisks $(\ast)$ are the results of
simulated ``observations''. The solid line is the prediction for $\xi=0$, corresponding to a pure exponential
decay in time.  \label{fig:Results}}
\end{figure}

\begin{figure}[h]
\includegraphics[width=\columnwidth]{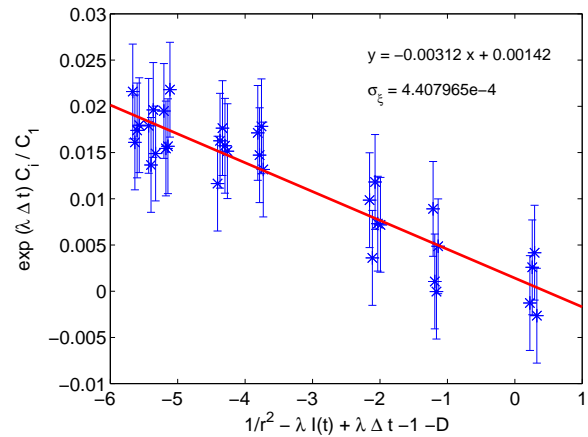}
\caption{Determination of $\xi$ from the simulated data. The vertical and horizontal axes correspond,
respectively, to the left-hand and right-hand sides of Eq. (\ref{SimRat}), where $C_i\equiv \dot{N_i}$, etc., 
and the solid line gives the
best fit to the simulated data. The inferred value of $\xi$ is given in Eq. (\ref{xiEqn}).  \label{fig:fit}}
\end{figure}

The results of our simulation are shown in Figs. \ref{fig:Results} and \ref{fig:fit}. In Fig. \ref{fig:Results}, the error bars represent the 
$1/\sqrt{N}$ errors in each point, and the asterisks ($\ast$) denote the result of a random measurement of the count rate ratio
within the $1/\sqrt{N}$ range. The solid line represents the expected results in the null-case ($\xi=0$),
corresponding to a pure exponential decay in time. Fitting the simulated data to Eq. (\ref{SimRat}), we find (see Fig. \ref{fig:fit}),
\begin{eqnarray}
\xi = (-3.1\pm{}0.4)\times10^{-3},
\label{xiEqn}
\end{eqnarray}
\noindent{}for the post-MOI data that we have generated. We note from Eq. (\ref{xiEqn}) that the central value of $\xi$ is in agreement with the input
value $\xi=-0.0031$ that we have used. More significantly, the uncertainty in $\xi$ is sufficiently small to
suggest that a non-zero value of $\xi$ near the nominal level that we have assumed might be detectable using
a more sophisticated analysis of the actual post-MOI MESSENGER data.

\section*{Acknowledgements}
The work of EF is supported in part by U.S. DOE contract No. DE-AC02-76ER071428.


%


\end{document}